\documentclass[10pt]{sig-alternate-05-2015}

\usepackage{graphicx}

\usepackage{amsmath}

\usepackage{url}

\usepackage{color}

\usepackage{xspace}

\usepackage{graphicx}

\usepackage{comment}

\usepackage{amssymb}

\usepackage{cleveref}

\usepackage{subcaption}

\usepackage[]{flushend}

\newcommand{\etal}{{et~al}.\@~}
\newcommand{\eg}{e.g.,\xspace}
\newcommand{\ie}{i.e.,\xspace}

\crefname{section}{\S}{\S\S}
\Crefname{section}{\S}{\S\S}
\crefname{subsection}{\S}{\S\S}
\Crefname{subsection}{\S}{\S\S}

\renewcommand\paragraph[1]{\smallskip\noindent\textbf{#1.}}

\newcommand{\name}{SDNsec\xspace}
\newcommand{\detour}{failover\xspace}

\begin{document}

\title{SDNsec: Forwarding Accountability\\ for the SDN Data Plane}

\author{
Takayuki Sasaki$^{\dagger}$, Christos Pappas$^{\ast}$, Taeho Lee$^{\ast}$, Torsten Hoefler$^{\ast}$, Adrian Perrig$^{\ast}$\\
\medskip
\normalsize
\begin{tabular}{cp{0.1cm}cp{0.1cm}c}
\\
$^{\dagger}$NEC Corporation&& $^{\ast}$ETH Z\"urich\\
t-sasaki@fb.jp.nec.com && 
\parbox{7cm}{\{pappasch, kthlee, htor, aperrig\}@inf.ethz.ch}
\end{tabular}
}

\maketitle

\begin{abstract}
SDN promises to make networks more flexible, programmable, and easier to
manage. Inherent security problems in SDN today, however, pose a threat to the
promised benefits. First, the network operator lacks tools to proactively
ensure that policies will be followed or to reactively inspect the behavior of
the network. Second, the distributed nature of state updates at the data plane
leads to inconsistent network behavior during reconfigurations. Third, the
large flow space makes the data plane susceptible to state exhaustion attacks.

This paper presents \name, an SDN security extension that provides forwarding
accountability for the SDN data plane. Forwarding rules are encoded in the
packet, ensuring consistent network behavior during reconfigurations and
limiting state exhaustion attacks due to table lookups. Symmetric-key
cryptography is used to protect the integrity of the forwarding rules and
enforce them at each switch. A complementary path validation mechanism allows
the controller to reactively examine the actual path taken by the packets.
Furthermore, we present mechanisms for secure link-failure recovery and
multicast/broadcast forwarding.

\end{abstract}

\section{Introduction}
\label{sec:intro}
Software Defined Networking (SDN) and its current realization --
OpenFlow~\cite{openflow} -- promise to revolutionize networking by centralizing
network administration and eliminating vendor lock-in. Rapid service
deployment, simplified network management, and reduced operational costs are
some of the promised benefits. Furthermore, SDN serves as a building block to
mitigate network security issues~\cite{fresco,veriflow,fortnox}.
Ironically, though, security of SDN itself is a neglected issue.

SDN is rife with vulnerabilities at the data plane. Compromised
switches~\cite{syn_knock,juniper_bug,nsa_backdoor} can redirect traffic over
unauthorized paths to perform eavesdropping, man-in-the-middle attacks, or to
bypass security middleboxes~\cite{detectswitch}.  Furthermore, they can disrupt
availability by launching state exhaustion attacks against other
switches~\cite{detectswitch,kloti2013,spooknet} or by simply dropping packets.
In addition, next generation botnets, consisting of compromised hosts and
switches, could unleash an unprecedented firepower against their victims. There
are latent vulnerabilities in SDN today that make these attacks feasible.

The first problem lies in the adversary model for the data plane:
all network devices are trusted to correctly follow the specified network policies.
Thus, the data plane lacks accountability mechanisms to verify that forwarding rules are
correctly applied. Specifically, it does not provide guarantees that the 
policies will be followed (\textit{enforcement}) nor proof that policies have
not been violated (\textit{validation}). Once one or more switches get
compromised, forwarding policies can be violated without getting caught
by other switches or the controller.

Another problem is the lack of consistency guarantees when the forwarding plane
is reconfigured~\cite{netupdate}. During reconfigurations, packets can follow paths that
do not comply with policy, leading to link flooding or isolation violations in
multitenant environments. This is an inherent problem in distributed systems,
because the new policy is correctly applied only after all affected switches
have been updated. However, an attacker can exploit the problem by
forcing reconfigurations through a compromised switch.

Our goal is to build an SDN security extension which ensures that the operator's
policies are correctly applied at the data plane through forwarding
accountability mechanisms.  That is, the extension should ensure consistent
policy updates, enforce network paths, and provide a means for operators to
reactively inspect how traffic has been forwarded.

There are only a few proposals dealing with SDN data-plane security.  A recent
security analysis of OpenFlow~\cite{kloti2013} proposes simple patch solutions
(rate limiting, event filtering, and packet dropping) to counter resource
exhaustion attacks. SANE~\cite{sane}, a pre-SDN era proposal, proposes a
security architecture to protect enterprise networks from malicious switches.
However, it lacks a validation mechanism to ensure that a path was indeed
followed; failure recovery is pushed to the end hosts.
Another class of proposals checks for policy violations by examining certain
network invariants; checking can be performed in real time during network
reconfigurations~\cite{veriflow,hsa,netplumber} or by explicitly requesting the
state of the data plane~\cite{anteater}.

\paragraph{Contributions}
This paper proposes an SDN security extension, \name, to achieve forwarding
accountability for the SDN data plane. Consistent
updates, path enforcement, and path validation are achieved through additional
information carried in the packets. Cryptographic markings computed by the
controller and verified by the switches construct a path enforcement mechanism;
and cryptographic markings computed by the switches and verified by the
controller construct a path validation mechanism. Furthermore, we describe
mechanisms for secure failure recovery.
Finally, we implement the \name data plane on software switches and
show that state exhaustion attacks are confined to the edge of the network.

\section{Problem Description}
\label{sec:probdef}
We consider a typical SDN network with a forwarding plane that implements
the operator's network policies through a logically centralized controller. 
Network policies of the operator dictate which flows are authorized to access
the network and which paths are authorized to forward traffic for the corresponding
flows. 

Our goal is to design an extension that makes a best-effort
attempt to enforce network policies at the forwarding plane, and to detect and
inform the controller in case of policy violations. 

\subsection{Adversary Model}
\label{subsec:adversary}

\begin{figure}[t]
	\centering
	\includegraphics[width=.8\columnwidth]{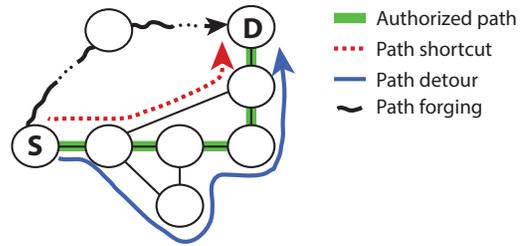}
	\caption{Forms of path deviation attacks that do not follow the authorized path from S to D.}
	\label{fig:path_deviation}
\end{figure}

The goal of the attacker is to subvert the network policies of the operator
(e.g., by forwarding traffic over unauthorized paths) or to disrupt the
communication between end hosts. To this end, we consider the following
attacks:

\paragraph{Path deviation} A switch causes packets of a flow to be
forwarded over a path that has not been authorized for the specific flow.
This attack can take the following forms (Figure~\ref{fig:path_deviation}):
\begin{itemize}
	\item \paragraph{Path detour} A switch redirects a packet to deviate 
	from the original path, but later the packet returns to the correct
	next-hop downstream switch.

	\item \paragraph{Path forging} A switch redirects a packet to deviate
	from the original path, but the packet does not return to a downstream
	switch of the original path.

	\item \paragraph{Path shortcut} A switch redirects a packet and skips
	other switches on the path; the packet is forwarded only by a subset
	of the intended switches.
\end{itemize}

\paragraph{Packet replay} A switch replays packet(s) to flood a host or 
another switch.

\paragraph{Denial-of-Service} We consider state 
exhaustion attacks against switches, which disrupt communication of
end hosts.

We consider an adversary that can compromise infrastructure components and
hosts, and can exploit protocol vulnerabilities. Furthermore, compromised
components are allowed to collude.

We do not consider payload modification attempts by
switches, as hosts do not trust the network and use end-to-end integrity checks
to detect any unauthorized changes. In addition, controller security is out of
the scope of this paper, since our goal is to enforce the controller policies
at the forwarding plane.

\subsection{Assumptions}
We make the following assumptions:
\begin{itemize}
\item Cryptographic primitives are secure, \ie hash functions cannot
be inverted, signatures cannot be forged, and encryptions cannot be broken. 

\item The communication channel between the controller and benign
switches is secure (\eg TLS can be used, as in OpenFlow~\cite{openflow}).

\item End hosts are authenticated to the controller and cannot spoof their
identity (\eg port-based Network Access Control can be used~\cite{8021X}).

\end{itemize}

\section{Overview}
\label{sec:overview}
In \name, the controller computes network paths and the corresponding
forwarding information. The switches at the edge of the network receive this
forwarding information over a secure channel and embed it into packets that
enter the network. Switches at the core of the network forward packets
according to the forwarding information carried in the packets; and the last
switch on the path removes the embedded information before forwarding the
packet to the destination. Figure~\ref{fig:fwplane} shows the network model for \name.
We stress that end hosts do not perform any additional functionality (\eg
communicate with the controller), \ie the network stack of the hosts is 
unmodified.

We describe and justify our main design decisions and present an overview of
the control and data plane.

\subsection{Central Ideas}
\label{ssec:key_concepts}

We identify three main problems that undermine network policies in today's SDN
networks and describe our corresponding design decisions.

\paragraph{Consistent Updates} In SDN, the distributed nature of updating the
forwarding plane can cause inconsistencies among switches. Specifically, a new
policy is correctly applied only after all affected switches have been
reconfigured; however, during state changes the forwarding behavior may be
ill-defined. Although solutions have been proposed to counter this
problem~\cite{Mahajan:2013, Dionysus}, they require coordination between the
controller and \textit{all} the involved switches in order to perform the updates.

In \name, packets encode the forwarding information for the intended path. This
approach guarantees that once a packet enters the network, the path to be
followed is fixed and cannot change under normal operation (\ie without link
failures).  Hence, a packet cannot encounter a mixture of old and new
forwarding policies, leading to inconsistent network behavior.  
Forwarding tables exist only at the entry and exit points of the network,
simplifying network reconfiguration: only the edge of the network must be updated
and coordination among all forwarding devices is not needed.

The packet overhead we have to pay for this approach provides additional benefits:
guaranteed loop freedom, since we eliminate asynchronous updates; and minimum
state requirements for switches, since forwarding tables are not needed in most
of the switches (see Section~\ref{ssec:fwd_plane}).  The lack of forwarding
tables confines the threat of state exhaustion attacks.

\paragraph{Path Enforcement} In SDN, the controller cannot obtain guarantees that the forwarding
policies will be followed, since the forwarding plane lacks enforcement mechanisms.
Ideally, when a switch forwards packets out of the wrong port, the next-hop switch 
detects the violation and drops the packet.

We incorporate a security mechanism that protects the integrity of the
forwarding information in order to detect deviations from the intended path and
drop the traffic. However, this mechanism by itself is insufficient to protect
from replaying forwarding information that has been authorized for other flows.

\paragraph{Path Validation} In SDN, the controller has no knowledge of the actual path that a packet
has taken due to the lack of path validation mechanisms. 

We design a reactive security mechanism that checks if the intended path was
followed. The combination of path enforcement and path validation provides protection
against strong colluding adversaries.

\subsection{Controller}
The controller consists of two main components: a path computation component (PCC) 
and a path validation component (PVC). Furthermore, the controller generates and shares
a secret key with every switch at the data plane; the shared key is communicated
over the secure communication channel between them.

\subsubsection{Path Computation Component}
\label{sssec:pcc}
The PCC computes the forwarding information for paths that are authorized for
communication.  Specifically, for each flow that is generated, a path is
computed. We do not impose restrictions on the flow specification; for
interoperability with existing deployments, we adopt the 13-tuple flow
specification of OpenFlow~\cite{of15}.

The computed forwarding information for a flow is embedded in every packet of
the flow.  For each switch on the path, the PCC calculates the egress interface
that the packet should be forwarded on\footnote{We assume a unique numbering assignment
for the ports of a switch.}. Hence, the ordered list of interfaces
specifies the end-to-end path that the packets should follow. Furthermore, each
flow and its corresponding path is associated with an expiration time
($\mathit{ExpTime}$) and a flow identifier ($\mathit{FlowID}$). The expiration
time denotes the time at which the flow becomes invalid, and the flow
identifier is used to optimize flow monitoring in the network (Section~\ref{ssec:monitoring}).

\begin{figure}
	\centering
	\includegraphics[width=.9\columnwidth]{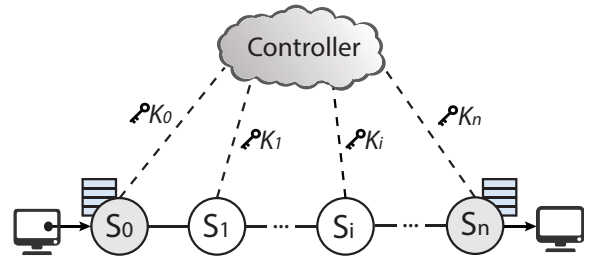}
	\caption{The \name network model: The ingress and egress switches store
	forwarding tables; and the controller has a shared secret with every switch 
	at the data plane.}
	\label{fig:fwplane}
\end{figure}

Furthermore, the forwarding information contains cryptographic primitives that
realize path enforcement. Each forwarding entry ($\mathit{FE}(S_i)$) for switch
$S_i$ contains a Message Authentication Code (MAC) that is computed over the
egress interface of the switch ($\textit{egr}(S_i)$), the flow information
($\textit{ExpTime}$ and $\textit{FlowID}$), and the forwarding entry of the
previous switch ($\textit{FE}(S_{i-1})$); the MAC is computed with the shared
key ($K_i$) between the controller and the corresponding switch on the path.
Equation~\ref{eq:fe} and Figure~\ref{fig:fwplane} illustrate how the forwarding
information is computed recursively for switch $S_i$ (for $1 \leq i
\leq n$).

\begin{equation}
\begin{aligned}
	&B =  \mathit{FlowID}~||~\mathit{ExpTime}\\
	&\mathit{FE}(S_i) = \mathit{egr}(S_i)~||~\mathit{MAC}(S_i)\\
	&\mathit{MAC}(S_i)=\mathit{MAC}_{K_i}(\mathit{egr}(S_i)~||~\mathit{FE}(S_{i-1})~||~B )\\
\end{aligned}
\label{eq:fe}
\end{equation}

Furthermore, a forwarding entry for switch $S_0$ is inserted into the packet to be used by $S_1$ for 
correct verification of its own forwarding information; $\mathit{FE}(S_0)$ is not used by the first-hop
switch and is computed as follows: $\mathit{FE}(S_0) = B$.

\subsubsection{Path Validation Component}
\label{sssec:pvc}
The PVC is a reactive security mechanism that provides feedback/information about the path
that a packet has taken. The controller can then detect attacks that have bypassed path enforcement
and reconfigure the network accordingly. Path validation is achieved through two mechanisms:
a path validation field in the packet and flow monitoring.

Each switch embeds a proof in every packet that it has indeed forwarded the packet. Hence, the collective
proof from all on-path switches forms a trace for the path that the packet has taken. The controller
can instruct any switch to report packet headers and thus inspect the path that was taken.

The path validation field of a switch ($\mathit{PVF}(S_i)$) contains a MAC that
is computed over the PVF of the previous switch ($\mathit{PVF}(S_{i-1}$)), flow
related information ($\mathit{FlowID}$), and a sequence
number ($\mathit{SeqNo}$).  The SeqNo is used to construct mutable information
per packet, ensuring different PVF values for different packets; this detects
replay attacks of the PVFs. The MAC is computed with the shared key between the
switch and the controller\footnote{For ease of exposition, the MAC of the PVF
is computed with the same key as the MAC of the FE. In a real deployment,
these two keys would be different.}. Equation~\ref{eq:pvf} shows how the PVF is computed:

\begin{equation}
\begin{aligned}
	& C = \mathit{FlowID}~||~\mathit{SeqNo}\\
	& \mathit{PVF}(S_0) = \mathit{MAC}_{K_0}(C)\\
	&\mathit{PVF}(S_i) = \mathit{MAC}_{K_i}(\mathit{PVF}(S_{i-1})~||~C), 1 \leq i \leq n\\
\end{aligned}
\label{eq:pvf}
\end{equation}

Given the $\mathit{FlowID}$ and $\mathit{PVF}$ in the packet header, the
controller can detect path deviations. The controller knows the path for the
given flow, and thus the keys of the switches on the path. Thus, the controller
can recompute the correct value for the PVF and compare it with the reported one.
However, this mechanism cannot detect dishonest switches that do not report all
packet headers when requested.

Monitoring and flow statistics are additional mechanisms to detect false
reporting.\footnote{Monitoring is an essential tool for other crucial tasks as
well (\eg traffic engineering).} The controller can instruct arbitrary switches
to monitor specific flows and obtain their packet counters. Inconsistent packet
reports indicate potential misbehavior and further investigation is required.
For instance, if all switches after a certain point on the path report a lower
packet count, then packets were possibly dropped. However, if only a switch in
the middle of the path reports fewer packets, it indicates a dishonest report.
The controller combines flow monitoring with the PVF in the packet headers to
detect policy violations.

\subsection{Data Plane}
\label{ssec:fwd_plane}

The data plane of \name consists of edge and core switches
(Figure~\ref{fig:fwplane}). Edge switches (shaded circles) operate at the
edge of the network and serve as the entry and exit points to the network.
Core switches operate in the middle of the network and forward packets based
on the forwarding information in the packets. 

\subsubsection{Edge Switches}
Edge switches are directly connected to network hosts and perform different
operations when acting as an entry point (ingress switch) and when acting as an
exit point (egress switch).  Edge switches, as opposed to core switches, have
flow tables in order to forward packets.

\paragraph{Ingress Switch}
An ingress switch receives packets from source hosts and uses a forwarding
table to look up the list of forwarding entries for a specific flow. In case of
a lookup failure, the switch consults the controller and obtains the corresponding
forwarding information.

Next, the switch creates a packet header and inscribes the forwarding information 
in it. Furthermore, for every packet of a flow, the switch inscribes a sequence
number to enable replay detection of the PVF. Finally, the switch 
inscribes $\mathit{PVF(S_0)}$, and forwards the packet to the next switch.

\paragraph{Egress Switch}
An egress switch receives packets from a core switch and forwards them to the
destination. To forward a packet, the egress switch uses a forwarding
table in the same way as the ingress switch. 

Having a forwarding table at the egress switch is a design decision that limits
the size of forwarding tables at ingress switches. It allows rule aggregation
at ingress switches at the granularity of an egress switch. Without a
forwarding table at the egress switch, a separate flow rule for every egress
port of an egress switch would be needed.  The egress switch has the egress
interface encoded in its FE, but it does not consider it when forwarding the
packet; the FE is still used to verify the correct operation of the previous
hop. 

Upon packet reception, the switch removes the additional packet header and 
forwards the packet to the destination. If requested, it reports the packet
header, together with its PVF to the controller.

\subsubsection{Core Switches}
Core switches operate in the middle of the network and perform minimal 
operations per packet. They verify the integrity of their corresponding forwarding entry and forward
the packet out of the specified interface. In case of a verification failure,
they drop the packet and notify the controller.

Furthermore, each core switch stores a list of failover paths that are used
in case of a link failure (Section~\ref{ssec:link_failure}) and keeps
state only for multicast/broadcast traffic (Section~\ref{ssec:mcast}) and 
flow monitoring (Section~\ref{ssec:monitoring}).

\section{Details}
\label{sec:details}
First, we present the \name packet header. Then, we describe link-failure
recovery, multicast/broadcast forwarding, and monitoring.

\subsection{\name Packet Header}
\label{ssec:pkt_header}
\begin{figure}[!t]
\centering
\includegraphics[width=1.\columnwidth]{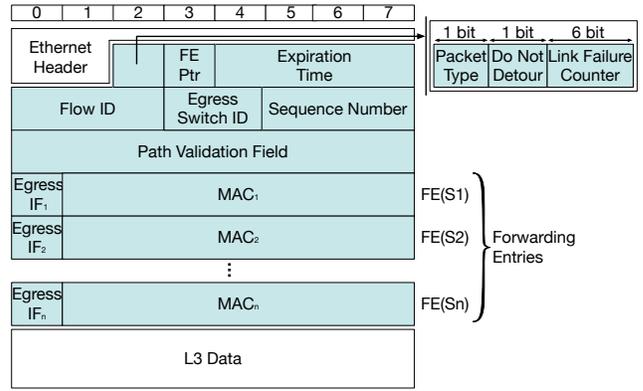}
\caption{\name packet header for unicast traffic.}
\label{fig:header_format}
\end{figure}

The packet header (Figure~\ref{fig:header_format}) encodes the forwarding
information (Equation~\ref{eq:fe}), the PVF (Equation~\ref{eq:pvf}), and
additional information that enables the switches to parse the header (\eg
a pointer to the correct forwarding entry). We present the packet-header
fields categorized by their use.

\subsubsection{Fields for Forwarding and Path Enforcement}

\begin{itemize}
    \item{\textbf{Packet Type(PktType)}: PktType indicates whether the packet
is a multicast/broadcast or a unicast packet. A single bit is used as a boolean
flag to indicate the packet type.}

    \item{\textbf{FE Ptr}: A pointer that points to the FE that a switch on the
path must examine. During packet processing, each switch increments the pointer so that
the next-hop switch examines the correct FE. One byte is allocated for the FE Ptr, 
which means that \name can support up to 255 switches for a single path. This upper bound
does not raise practical considerations even for large topologies, since the network
diameter is typically much shorter.}

    \item{\textbf{Expiration Time (ExpTime)}: ExpTime indicates the time after
which the flow becomes invalid. Switches discard packets with expired
forwarding information. ExpTime is expressed at the granularity of one second, and
the four bytes can express up to 136 years.}

    \item{\textbf{Forwarding Entry (FE)}: A FE for switch $S_i$ consists of the egress interface
of switch $S_i$ ($\textit{egr}(S_i)$) and the MAC ($\textit{MAC}(S_i)$)
that protects the integrity of the partial path that leads up to the switch $S_i$.
One byte is used for $\textit{egr}(S_i)$ allowing each switch to have up to 255
interfaces; and 7 bytes are used for $\textit{MAC}(S_i)$. In
Section~\ref{ssec:mac_len}, we justify why a 7-byte MAC is sufficient to ensure
path integrity.}

\end{itemize}

\subsubsection{Fields for Path Validation}

\begin{itemize}

    \item{\textbf{Path Validation Field (PVF)}: Each switch that forwards the
packet inserts a cryptographic marking on the PVF according to
Equation~\ref{eq:pvf}, and the controller uses the PVF for path validation.
\name reserves 8 bytes for PVF, and in Section~\ref{ssec:mac_len}, we justify that 8 bytes
provide sufficient protection against attacks.}

    \item{\textbf{Sequence Number (SeqNo)}: The ingress switch inserts a
monotonically increasing packet counter in every packet it forwards.
Specifically, a separate counter is kept for every flow entry at the ingress
switch.  The SeqNo is used to randomize the PVF and to detect replay attacks
against the Path Validation mechanism, in which a malicious switch replays
valid PVFs to validate a rogue path. The 24-bit sequence number can identify more
than 16 million unique packets for a given flow. For the average packet size of
850 bytes in data centers~\cite{datac_traffic_ccr}, the 24 bits suffice for a
flow size of 13~GB; Benson et al. report that the maximum flow size is less
than 100~MB for the 10 data centers studied~\cite{datac_traffic}. Hence, it is
highly unlikely that the sequence number wraps around. Even if the sequence
number wraps around, under normal operation the same values would appear
a few times, whereas in an attack scenario typically a high repetition rate 
of certain values would be observed.}

    \item{\textbf{Flow ID (FlowID)}: FlowID is an integer that uniquely
identifies a flow. FlowIDs are used to index flow information,
enabling \name entities (controller and switches) to efficiently search for
flow information; the 3 bytes can index over 16 million flows. The
active flows in four data centers, as observed from 7 switches in the network,
do not exceed 100,000~\cite{datac_traffic}.}

\end{itemize}

\subsubsection{Fields for Link-Failure Recovery}
\begin{itemize}

    \item{\textbf{Link Failure Counter (LFC)}: LFC indicates the number of
failed links that a packet has encountered throughout its journey towards the
destination. \name reserves 6 bits for LFC, which means that up to 63 link failures 
can be supported (see Section~\ref{ssec:link_failure}).}

    \item{\textbf{Egress Switch ID (EgressID)}: The EgressID identifies the
egress switch of a packet. Although FEs in the packet dictate the sequence of
switches that a packet traverses, the core switches on the path cannot
determine the egress switch (except for the penultimate core switch) from the
FEs. However, the egress switch information is necessary when a core switch
suffers a link failure and needs to determine an alternate path to the egress
switch. To this end, the \name header contains the EgressID. With 2 bytes, it
is possible to uniquely identify 65,536 switches, which is sufficient even for
large data centers.}

\end{itemize}

\subsection{Link-Failure Recovery}
\label{ssec:link_failure}
The design decision that packets encode the forwarding information for the intended
path makes link-failure recovery challenging: the intended path for packets that
are already in the network is no longer valid. Dropping all ill-fated packets does not
compromise the security guarantees, but degrades network availability until the
controller reconfigures the network.

We design a temporary solution to account for the ill-fated packets until a new
path is specified at the corresponding ingress switches or until the failure is
fixed. Furthermore, the temporary solution must satisfy the three requirements for
\name. First, it must ensure update consistency, \ie only one temporary policy 
must be used on one packet for one link failure. Second, it must provide path
enforcement, \ie deviations from the intended temporary path should lead to
packet dropping from a benign switch. Third, it must enable path validation,
\ie the controller must be able to verify the path -- including the switches
of the temporary policy -- that a packet has taken.

Our recovery mechanism uses a failover path. A failover path is a
temporary path that detours around the failed link and leads to the same egress
switch as the original path. The forwarding information of the failover path is
encoded in the packet as described in Equation~\ref{eq:fe}. That is, the
failover path contains the list of egress interfaces of the switches that are
on the detour path; the integrity of the list is protected with MACs that are
computed with the corresponding keys of these switches. When a link failure is
detected, the switch inserts the appropriate pre-computed failover path into
the packet and forwards the packet to the appropriate next hop, as specified by
the failover path. Each switch on the failover path updates the PVF as it would
do for a normal path. Since the failover path is constructed identically to the
original path, the forwarding procedure (Section~\ref{ssec:fwd_plane}) needs
only minor modifications (Section~\ref{sssec:mod_fwd}).

This solution satisfies the mentioned requirements. First, update consistency
is satisfied since the forwarding information of the failover path is encoded
in the \name header. Second, the authenticated forwarding information provides
path enforcement. Third, the controller can perform path validation -- including
the failover path -- with minor changes.

One shortcoming of the recovery mechanism is the requirement to store state at core 
switches for the pre-computed failover paths. To balance the tradeoff between 
fine-grained control and state requirements, core switches store per-egress-switch
failover paths. Alternative solutions could store per-flow or per-link failover paths.
Per flow failover paths provide very fine-grained control, since the operator can
exactly specify the path for a flow in case of a failure. However, core switches would 
have to store failover paths for every flow they serve. Per-link failover paths
minimize the state requirements, but provide minimal control to the operator\footnote{
Per-link failover paths would detour the ill-fated packets to the next-hop switch
of the original path, but over another temporary path.}. Furthermore, a path to the
egress switch might exist, even if a path around the failed link to the next-hop
switch does not.

Storing per-egress-switch failover paths may not satisfy the strict isolation
requirements for certain flows. For example, the failover path to the egress
switch may traverse an area of the network that should be avoided for specific
flows. To this end, we define a \textit{do not detour flag}. If set, the switch
drops the packet instead of using the failover path. In other words, the flag
indicates if security or availability prevails in the face of a link failure.
Note that failover paths are temporary fixes to increase availability, while the
controller computes a permanent solution to respond to the failure.

\subsubsection{Forwarding with Failover Paths}
\label{sssec:mod_fwd}

\begin{figure}
    \centering
    \includegraphics[width=1.\columnwidth]{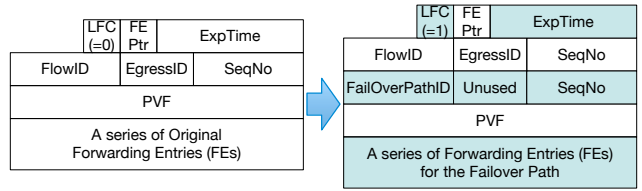}
    \caption{Modifications to the \name packet header for link-failure recovery;
	additional and modified fields are highlighted.}
    \label{fig:backup_path_header}
\end{figure}

\paragraph{Packet Header} Figure~\ref{fig:backup_path_header} shows how a
switch changes the packet header of an ill-fated packet when a failover path is
used. The FEs of the original path are replaced with those of the failover
path. Furthermore, the switch changes the expiration time field with
$ExpTime_{FailoverPath}$ and appends the information of the failover path (\ie
$FailoverPathID$, $SeqNo$) below that of the original path. Hence, the packet
contains the flow information of the original and the failover paths followed
by the FEs of the failover path.

Then, the switch resets \textit{FE Ptr} to one so that the next-hop switch on
the \detour path can correctly determine the FE that it needs to examine.

Lastly, the switch increments the LFC by one to indicate that a link-failure
has occurred. The LFC field counts the number of failover paths that a packet
has taken and enables multiple link failures to be handled without additional
complexity.

\paragraph{Forwarding Procedure} Three changes are made to the forwarding
procedure to accommodate link failures. First, since additional forwarding
information is inserted into the packet if there is a detour, a switch
identifies the correct FE by computing the following byte offset from the
beginning of the \name packet header: $6+(LFC+2)*8+FE Ptr*8$ bytes. Second,
when computing the PVF, the switch uses Equation~\ref{eq:mod_pvf} if there is a
detour. $FailOverPathID$ is determined by looking at the FlowID field of the
most recent forwarding information, which is identified by taking the byte
offset of $6+LFC*8$ bytes. 

\begin{equation}
    \mathit{C=FailOverPathID~||~SeqNo}
    \label{eq:mod_pvf}
\end{equation}

\subsubsection{Path Validation}
\label{sssec:path_validation}
Path validation accounts for the switches on the original path and the failover path.
The controller obtains the switches of the path that the packet should have traversed
by referring to the FlowID field(s) of the forwarding information in the header. Then using
Equation~\ref{eq:pvf} for the original path and Equation~\ref{eq:mod_pvf} for
the failover path(s), the controller computes the expected PVF value and
compares it with the PVF value in the packet header.

\subsection{Multicast/Broadcast}
\label{ssec:mcast}

We describe our design for multicast/broadcast forwarding that adheres to the three
requirements for \name (update consistency, path enforcement, and path validation).
For simplicity, we refer to multicast/broadcast forwarding as multicast.

A strawman's solution for multicast is to leverage unicast forwarding:
the ingress switch replicates each packet of a multicast group and uses the unicast
forwarding mechanism to send it to every egress switch that is on the path of a 
receiving host. This approach comes with two benefits: all three requirements are
satisfied; and the unicast forwarding mechanism can be used without modifications.
However, this solution is inefficient with respect to bandwidth overhead.

An alternative approach to implement multicast is to encode the multicast tree
in the packet.  Bloom filters can be used to efficiently encode the links of
the tree~\cite{lipsin}.  For each link, the switch checks if the bloom filter
returns a positive answer and forwards the packet along the corresponding
links. However, the false positives of Bloom filters become a limitation:
loops can be formed; and more importantly, forwarding a packet to an incorrect
switch violates network isolation.

We thus adopt a stateful multicast distribution tree to forward multicast
traffic. To implement forwarding along the
specified tree, the forwarding decisions are stored in forwarding tables at
switches. A multicast tree is represented by a two-tuple: an integer that
identifies the tree (\textit{TreeID}) and an expiration time (\textit{ExpTime})
that indicates when the tree becomes invalid.

\begin{figure}
\centering
\includegraphics[width=.9\columnwidth]{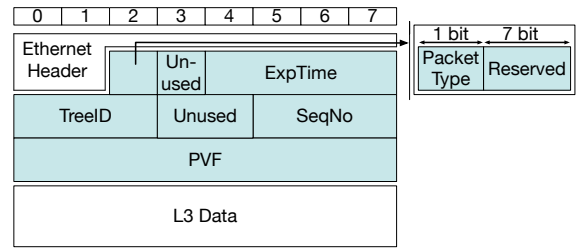}
\caption{\name packet header for multicast traffic.}
\label{fig:multicast_header}
\end{figure}

The controller computes a multicast tree and assigns it a unique \textit{TreeID}.
Then, it sends to each switch on the tree the two-tuple and the list of egress
interfaces. Upon receiving a multicast packet, the ingress switch determines the
correct multicast tree (based on the packet's information) and inserts the 
\textit{TreeID}, \textit{ExpTime}, and a sequence number (\textit{SeqNo}) in the
packet (Figure~\ref{fig:multicast_header}). Each core switch that receives a
multicast packet, looks up the forwarding information based on the \textit{TreeID} 
in the packet and forwards it according to the list of specified interfaces.

The main challenge with this stateful approach is policy consistency, \ie ensuring that a 
packet is not forwarded by two different versions of a multicast tree. To this end,
we require that a tree is never updated, instead a new tree is created. However, this
alone is not sufficient to guarantee drop freedom: if the ingress switch forwards 
packets of a newly created tree while core switches are being updated, then the
packets with the new \textit{TreeID} may get dropped by core switches.

To solve the problem, we add a safe-guard when switches are updated with a new
multicast tree: an ingress switch is not allowed to use the new tree, \ie insert
the \textit{TreeID} into incoming packets, until all other switches on the tree
(core and egress switches) have been updated with the new tree information. Ingress
switches can use the new tree only after an explicit notification by the controller.

Path enforcement is implemented implicitly, since only switches on the
multicast tree learn the two-tuple information of the tree; packets with
unknown \textit{TreeIDs} are dropped. Hence, if a malicious switch incorrectly
forwards a packet to an incorrect next-hop switch, then the switch will drop
the packet. Tampering with the \textit{TreeID} in the packet is detected through
path validation.

The path validation information for multicast is similar to unicast forwarding.
Each switch on the tree computes a MAC for the PVF using its shared key with the
controller. The only difference is that the \textit{TreeID}, instead of the
\textit{FlowID}, becomes an input
to the MAC (Equation~\ref{eq:treeidpvf}).

\begin{equation}
	C = \mathit{TreeID~||~SeqNo}
	\label{eq:treeidpvf}
\end{equation}

\newpage
\subsection{Monitoring}
\label{ssec:monitoring}
Network monitoring is an essential tool for traffic engineering and security
auditing. For instance, network operators can steer traffic away from traffic
hot spots or identify switches that drop packets.

In \name, monitoring is performed at the granularity of a flow, similar to 
OpenFlow. Switches maintain a monitoring table that stores packet counters 
for the flows that they serve. Specifically, ingress switches have flow tables
to look up the FEs, hence, an additional field is required for packet counters.
Core switches need an additional data structure to accommodate flow statistics.

Designing monitoring for the core network is based on two principles.
First, to prevent state exhaustion attacks the controller instructs switches
explicitly which flows they should monitor. Since switches do not monitor all
flows, an attacker cannot generate flows randomly to exhaust the monitoring table.
Second, to minimize the impact of monitoring on forwarding performance, we use
an exact match lookup table: the \textit{FlowID} in the packet header serves
as the key to the entry. Avoiding more heavyweight lookups (\eg longest prefix
matching) that require multiple memory accesses and often linear search operations
(\eg flow-table lookups in software switches) mitigates attacks that target
the computational complexity of the lookup procedure.

\section{Security Analysis}
\label{sec:seceval}
We start by justifying our design choice of short MACs, and then we
describe how \name protects from the attacks described in
Section~\ref{subsec:adversary}. 

\subsection{On the length of MACs}
\label{ssec:mac_len}
The path enforcement and path validation mechanisms require
MAC computations and verifications at every switch. We argue
that the length of the MACs -- 7 bytes for FEs and 8 bytes for 
the PVF -- is sufficient to provide the security guarantees we seek.

The main idea is that the secret keys used by other switches are
not known to the attacker, which means that an attacker can at
best randomly generate MACs without a way to check their validity.
Consequently, the attacker would have to inject an immense amount
of traffic even for a single valid FE ($2^{56}$ attempts are
required). Furthermore, to forge FEs for $n$ hops requires
$2^{56\cdot n}$ attempts, which becomes computationally infeasible
even for $n=2$. Hence, such traffic injection with incorrect MACs is
easily detectable.

\subsection{Path Deviation Attacks}
Path deviation attacks -- in which packets follow a path not
authorized by the controller -- can take different forms, as
described in Section~\ref{subsec:adversary}.

The security properties of chained MACs with respect to path validation have
been formalized and verified for a decentralized setting~\cite{mech_pathv}. The
centralized control in SDN simplifies key management, since the controller sets
up the shared symmetric keys with the switches; sophisticated key-establishment
protocols are not needed.  However, an important difference is
that we consider the ingress and egress switch -- not the hosts -- as the
source and destination, respectively. In Section~\ref{subsec:sec_con}, we
discuss the security implications of this decision.

Path enforcement is the first line of defense against path deviation
attacks. It prevents path forging and path detours from a malicious
switch that generates forged FEs. The next benign switch on the path
will drop the packet due to a MAC verification failure. However, a 
more sophisticated attacker can replay forwarding information of
other paths that it is part of, but which are not authorized for the
diverted flow.

Path validation is the second line of defense against path deviation attacks.
Since each switch is inscribing a MAC value in the packet, the packet carries
information about the presence or absence of switches on the path. The
controller can reactively inspect this information and obtain a guarantee about
the traversed switches and their order. \name provides this guarantee because
the attacker does not possess the secret keys of other switches.
Note that path validation also catches attacks from malicious ingress switches
that embed in the packets FEs of other flows. The controller knows the
forwarding information for every flow (based on the flow tuple) and can
detect the misbehavior. Changing the information that defines a flow would break
communication between the end hosts; Section~\ref{subsec:sec_con} discusses such
cases in more detail.

Furthermore, sequence numbers are used to prevent replay of the path validation
information. A malicious switch could replace the PVF value in a packet with a
value from a previously seen packet, obfuscating the actual path taken by the
packet to avoid being detected by the controller. The replay is detected
through a high repetition frequency of certain sequence numbers; under normal
operation each sequence number would appear at most a few times
(Section~\ref{ssec:pkt_header}).

The path enforcement and validation properties of \name can be compromised in
the case of multiple adjacent malicious switches. For example, if a malicious
on-path switch has multiple malicious adjacent switches (not on the path),
then the packets can be forwarded along the malicious path segment and back.
The on-path malicious switch can then reinject the packets
along the initial intended path; this attack cannot be detected, as pointed
out by prior work~\cite{mech_pathv}.

\subsection{Denial-of-Service}
Network devices typically store state (\eg forwarding tables) on 
fast memory (\eg SRAM), which is a limited resource. This becomes
the target of attackers by populating the memory with bogus data
that replaces legitimate information.

In \name, the state exhaustion attack vector is confined to the edge of the
network. Only edge switches keep forwarding tables and thus they are
susceptible to a state exhaustion attack by malicious hosts that originate
bogus flows. Core switches keep forwarding state only for broadcast/multicast
traffic, but these entries are preconfigured by the controller with the valid
tree IDs and, thus, cannot be populated with bogus entries. In
Section~\ref{ssec:state_ex}, we compare the performance between an edge switch
and a core switch under a state exhaustion attack.

Furthermore, each switch keeps state to monitor forwarded traffic at
the granularity of flows. An attacker could generate random flow IDs
in order to exhaust the monitoring table. This resource is protected
by having the switches monitor only flow IDs that the controller
mandates. Thus, the controller can securely adapt the resources
according to the device's capabilities.

\section{Implementation and Evaluation}
\label{sec:eval}
We implement the \name data-plane functionality on a software switch,
and evaluate performance on a commodity server machine. Furthermore,
we analyze the path validation and bandwidth overhead for the network.

\subsection{Software Switch Prototype}
To achieve high performance, our implementation leverages
the Data Plane Development Kit (DPDK)~\cite{dpdk} and the Intel AES-NI
instruction set~\cite{aesni}. DPDK is an open-source set of libraries and
drivers for packet processing in user space. DPDK comes with zero-copy Network
Interface Card (NIC) drivers that leverage polling to avoid unnecessary
interrupts. Intel AES-NI is an instruction set that uses hardware
cryptographic engines built into the CPU to speed up the AES block cipher.

To compute and verify the required MACs, we use the Cipher Block Chaining mode
(CBC-MAC) with AES as the block cipher. The input lengths to the MACs for a FE
and PVF are 15 and 14 bytes respectively. Note that for both cases the input
fits in one AES block (16 bytes) and that the input length is fixed and
independent of the path length\footnote{CBC-MAC is vulnerable when used for
variable-length messages}. Furthermore, we use 128-bit encryption keys and
truncate the output to the required number of bits (Section~\ref{ssec:pkt_header}).

Furthermore, we optimize forwarding in the following ways. First, we store four
FEs in different \texttt{xmm} registers (\texttt{xmm0-xmm3}) and issue four
encryption instructions with the preloaded round key (stored in \texttt{xmm4}).
Since each AES engine can simultaneously perform 4 AES operations, 
a switch can process four packets in parallel on each CPU core. The assembly code snippet is
given below:

\indent \texttt{\footnotesize aesenc xmm0,xmm4 //Round 1 for Packet 1}\\
\indent \texttt{\footnotesize aesenc xmm1,xmm4 //Round 1 for Packet 2}\\
\indent \texttt{\footnotesize aesenc xmm2,xmm4 //Round 1 for Packet 3}\\
\indent \texttt{\footnotesize aesenc xmm3,xmm4 //Round 1 for Packet 4}\\

Second, a dedicated CPU core is assigned to a NIC port and handles all the required
packet processing for the port. Each physical core has a dedicated AES-NI engine
and thus packets received on one port are served from the AES-NI engine of the
physical core assigned to that port.

Third, we create per-core data structures to avoid unnecessary cache misses.
Each NIC is linked with a receive queue and a transmit queue, and these queues
are assigned to a CPU core to handle the NIC's traffic. Furthermore, we load
balance traffic from one NIC over multiple cores, depending on the system's
hardware. For this purpose, we leverage Receiver Side Scaling (RSS)~\cite{rss} as
follows: each NIC is assigned multiple queues, and each queue can be handled by
another core. RSS is then used to distribute traffic among the queues of a
NIC.

Our implementation of the edge switch is based on the DPDK
vSwitch~\cite{dpdkvswitch}. The DPDK vSwitch is a fork of the open source
vSwitch~\cite{openvswitch} running on DPDK for better performance. Open vSwitch
is a multilayer switch that is used to build programmable networks and can run
within a hypervisor or as a standalone control stack for switching devices.
Edge switches in \name use the typical flow matching rules and forwarding
tables to forward a packet and therefore we chose to augment an existing
production quality solution. We augment the lookup table to store forwarding
information for a flow in addition to the output port. The ingress switch increases
the size of the packet header and inputs the additional information (FEs, sequence number,
and its PVF).

We implement core switches from scratch due to the minimal functionality they
perform. A core switch performs two MAC computations (it verifies its FE and
computes its PVF value), updates the flow's counters (if the flow is
monitored), and forwards the packet from the specified port.

\subsection{Packet Overhead}
\label{subsec:pkt_overhead}
The security properties of \name come at the cost of increased packet size.
For each packet, the ingress switch creates an additional packet header with
its size depending on the path length: 8 bytes/switch (including the 
egress switch) and a constant amount of 22 bytes/packet.

To put the packet overhead into context, we analyze two deployment scenarios
for \name: a data-center deployment and a research network deployment.
Furthermore, to evaluate the worst case for \name, we consider the diameter of
the network topologies, \ie the longest shortest path between any two nodes in
the network. We also evaluate the packet overhead for the average path length
in the research-network case.

\begin{table}[!b]
\small
\centering
\begin{tabular}{l|c|c|c|l}
                               & \multicolumn{3}{c}{Packet Size} &  \\
                               & 200 B          & 850 B         & 1400 B       &  \\ \hline\hline
\multicolumn{1}{c|}{Leaf-Spine} & 19.0\%      & 4.5\%      & 2.7\%     &  \\
3-Tier                         & 27.0\%      & 6.4\%      & 3.9\%     & 
\end{tabular}
\caption{Packet overhead for data center traffic patterns and topologies.}
\label{table:datacenter}
\end{table}

For the data-center case, we consider two common data center topologies: a
leaf-spine topology~\cite{leafspine} and a 3-tier topology (access,
aggregation, and core layer)~\cite{multitier}. The diameter for the leaf-spine
topology is 4 links (\ie 3 switches) and for the 3-tier topology 6 links (\ie 5
switches)\footnote{Our reported path lengths include the links between the hosts and
the switches.}. In addition, to relate the overhead to realistic data center
traffic, we use the findings of two studies: the average packet size in data
centers is 850 bytes~\cite{datac_traffic_ccr}, and packet sizes are
concentrated around the values of 200 and 1400 bytes~\cite{datac_traffic}.
Table~\ref{table:datacenter} shows the overhead for the different topologies
and path lengths.

\begin{table}[!t]
\small
\centering
\begin{tabular}{lcc|cc|cc}
                 & \multicolumn{2}{c|}{Trace 1}                               & \multicolumn{2}{c|}{Trace 2}                              & \multicolumn{2}{c}{Trace 3}                               \\
                 & 747 B                         & 463 B                         & 906 B                         & 1420 B                       & 691 B                         & 262 B                         \\\hline\hline
A  & 8.3\%                       & 13.4\%                     & 6.8\%                      & 4.4\%                     & 9.0\%                      & 23.7\%                     \\
D         & \multicolumn{1}{l}{12.6\%} & \multicolumn{1}{l|}{20.3\%} & \multicolumn{1}{l}{10.4\%} & \multicolumn{1}{l|}{6.6\%} & \multicolumn{1}{l}{14.0\%} & \multicolumn{1}{l}{35.9\%}
\end{tabular}
\caption{Packet overhead for the average path length (A) and the diameter (D) of the Internet2 topology and the mean and median
packet sizes from 3 CAIDA 1-hour packet traces.}
\label{table:internet2data}
\end{table}

For the research network deployment, we analyze the topology of the Internet2
network~\cite{internet2}, which is publicly available~\cite{internet2_noc}; we
consider only the 17 L3 and 34 L2 devices in the topology -- not the L1 optical
repeaters -- and find a diameter of 11 links (\ie 10 switches). Furthermore, for the 
Internet2 topology we calculate an average path length of 6.62 links (\ie 6 switches).
To relate the overhead to actual traffic, we analyze three 1-hour packet traces from
CAIDA~\cite{caida} and calculate the respective packet overhead for the mean
and median packet lengths.  (Table~\ref{table:internet2data}). 

Our results indicate a moderate packet overhead for the average path length in
Internet2 and a considerable packet overhead for the worst case (high path
lengths) in both deployment scenarios. This analysis provides an insight about
the price of security and robustness for policy enforcement and validation of
the SDN data plane. Furthermore, we observe that the packet overhead is more
significant for ISP topologies because they have typically longer paths than
data center networks:  data center networks are optimized with respect to
latency and cabling length leading to shorter path lengths. Novel data center
topologies demonstrate even shorter path lengths compared to the more common
topologies we analyzed~\cite{jellyfish}. This path-length optimization leads to
a lower packet overhead for a data center deployment of \name.

\subsection{Performance Evaluation} 

We compare the forwarding performance of edge and core switches with the DPDK
vSwitch under two scenarios: normal operation and a state exhaustion attack.

We run the \name software switch on a commodity server machine. The server has
a non-uniform memory access (NUMA) design with two Intel Xeon E5-2680 CPUs that
communicate over two QPI links.  Each NUMA node is equipped with four banks of
16~GB DD3 RAM.  Furthermore, the server has 2 dual-port 10~GbE NICs (PCIe
Gen2x8) providing a unidirectional capacity of 40~Gbps.

We utilize Spirent SPT-N4U-220 to generate traffic. We specify IPv4 as the
network-layer protocol, and we vary Ethernet packet sizes from 128 to 1500
bytes.\footnote{We exclude 64-byte packets because the minimum packet size in
the core of the network is higher because the additional information
in \name does not fit in the minimum-sized Ethernet packet.} For a given link
capacity, the packet size determines the packet rate and hence the load on the
switch.  For example, for 128-byte packets and one 10~GbE link, the maximum
packet rate is 8.45 Million packets per second (Mpps); for all 8 NIC ports it is
67.6~Mpps. These values are the physical limits and represent the theoretical peak
throughput.

Furthermore, for the \name edge switch and the DPDK vSwitch, we populate a flow
table with 64k entries; for the \name edge switch, the flow table holds forwarding
entries for a path with 5 switches. Flows are defined based on the destination MAC address
-- all other fields remain constant.

\subsubsection{Normal Operation}

\begin{figure*}[t]
\centering
\begin{subfigure}{.8\columnwidth}
    \centering
    \includegraphics[width=.8\columnwidth]{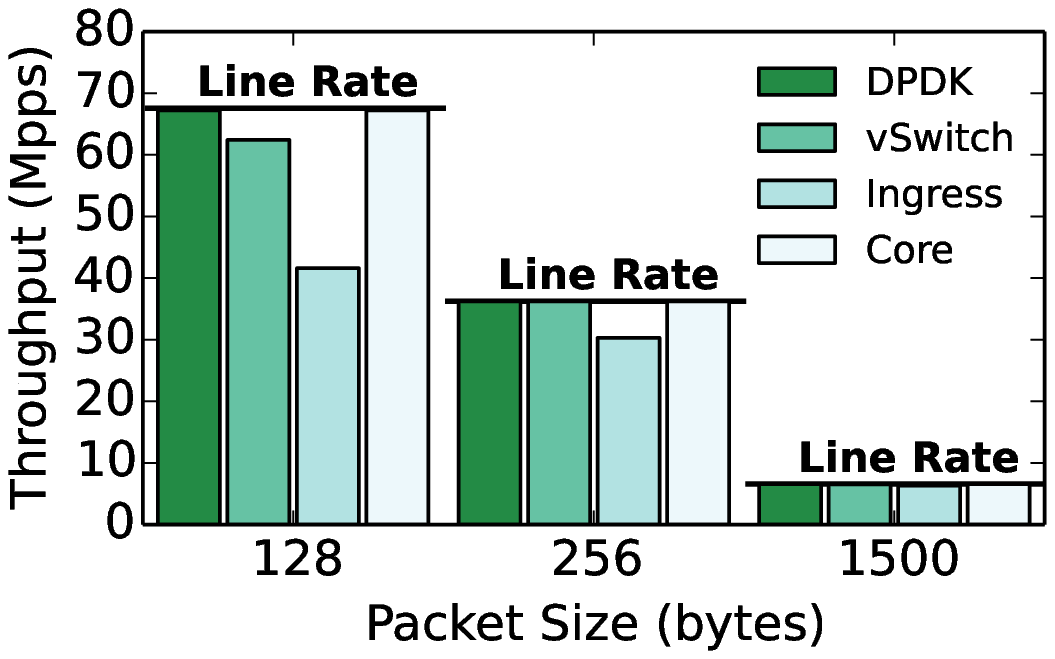}
    \caption{Throughput}
    \label{fig:norm_throughput}
\end{subfigure}
\hspace{40pt}
\begin{subfigure}{.8\columnwidth}
    \centering
    \includegraphics[width=.8\columnwidth]{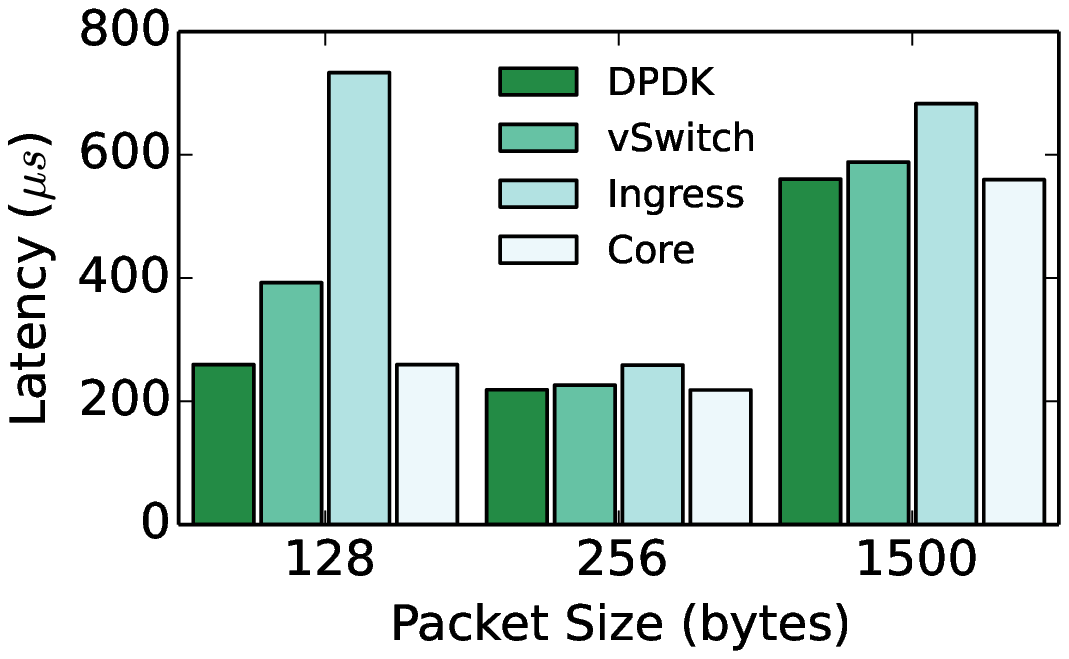}
    \caption{Latency}
    \label{fig:norm_latency}
\end{subfigure}
\caption{Switching performance under normal operation.}
\label{fig:ingress_latency}
\end{figure*}

For normal operation, we generate packets
with a destination MAC address in the range of the addresses stored in the flow
table of the switch. Figure~\ref{fig:norm_latency} shows the average latency
per packet, and Figure~\ref{fig:norm_throughput} shows the switching
performance for a 60-second measurement interval.

The ingress switch demonstrates a higher latency compared to DPDK vSwitch
because the \name header must be added to every packet: the packet size
increases and the longer entries in the lookup table cause additional cache
misses that increase latency. Furthermore, the latency of the core switch is
the same as the DPDK baseline latency, demonstrating the minimal processing
overhead at the core switches.

We observe a considerable performance decrease for the ingress switch compared
to the DPDK vSwitch. This decrease is a side-effect of the packet overhead
(Section~\ref{subsec:pkt_overhead}): the outgoing traffic volume of an ingress
switch is higher than the incoming volume. Thus, when the incoming links are
fully utilized, packets get dropped and the throughput is lower (assuming that
the aggregate ingress and egress capacity of the switch is the same). This
comparison captures the effect of packet overhead and not the processing
overhead. In contrast to the ingress switch, the core switch outperforms the
other switches and achieves the baseline performance for all packet sizes.

Our experiments under normal operation demonstrate a performance decrease at the
edge of the network, however, the core of the network can handle significantly
more traffic, compared to today's SDN realization.

\subsubsection{State Exhaustion Attack}
\label{ssec:state_ex}

\begin{figure*}[t]
\centering
\begin{subfigure}{.8\columnwidth}
    \centering
    \includegraphics[width=.8\columnwidth]{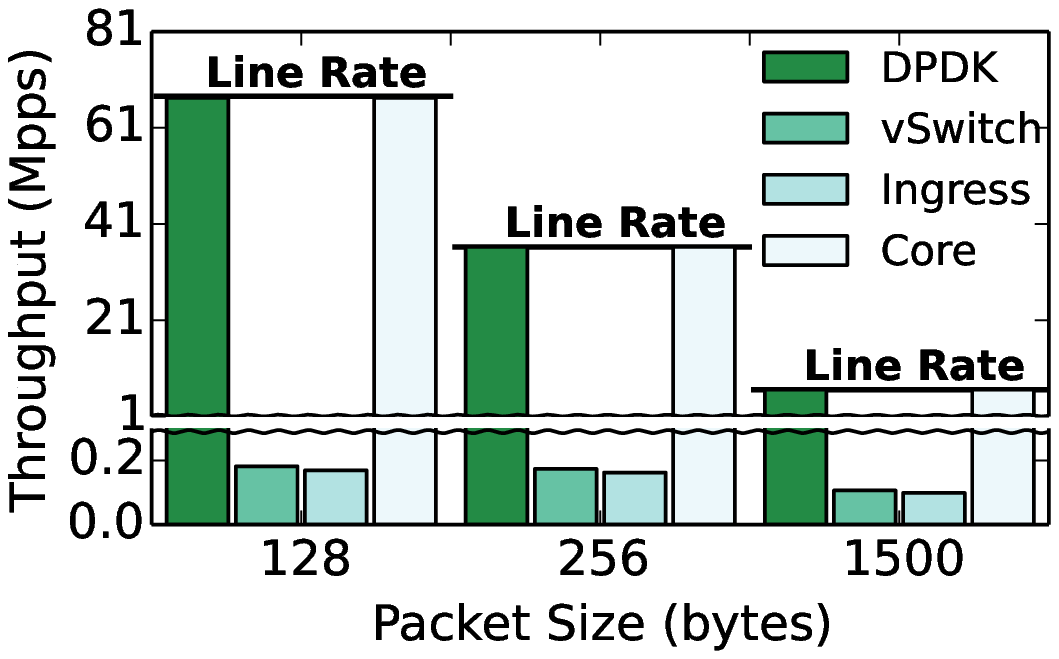}
    \caption{Throughput}
    \label{fig:se_throughput}
\end{subfigure}
\hspace{40pt} 
\begin{subfigure}{.8\columnwidth}
    \centering
    \includegraphics[width=.8\columnwidth]{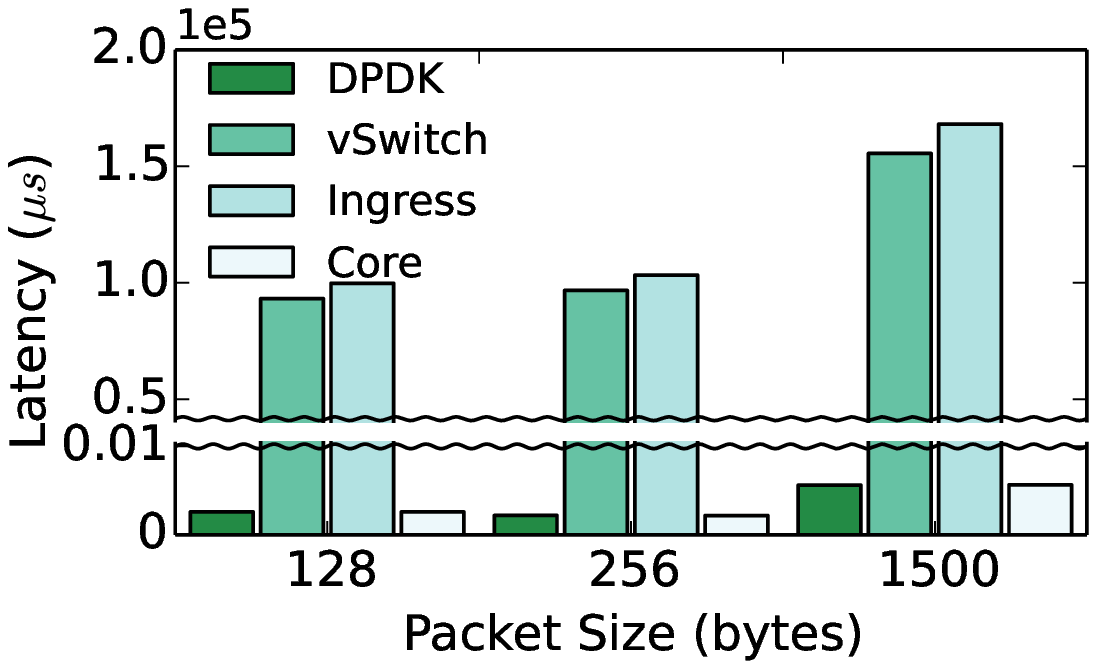}
    \caption{Latency}
    \label{fig:se_latency}
\end{subfigure}
\caption{Switching performance under state exhaustion attack.}
\label{fig:se_perf}
\end{figure*}

To analyze the switching performance of the switch under a state exhaustion
attack, we generate traffic with random destination MAC addresses. The
destination addresses are randomly drawn from a pool of $2^{32}$ ($\sim$4
billion) addresses to prevent the switches from performing any optimization,
such as caching flow information. Figure~\ref{fig:se_perf} shows the switching
performances.

We observe a considerable decrease (\ie over 100 times slower than the DPDK
baseline) in throughput for both the DPDK vSwitch and the ingress switch
(Figure~\ref{fig:se_throughput}). This decrease is due to cache misses when
performing flow table lookups--the switches are forced to perform memory
lookups, which are considerably slower than cache lookups, in order to determine the
forwarding information to process the incoming packets. The latency plot in
Figure~\ref{fig:se_latency} tells a similar story: both the DPDK vSwitch and
the ingress switch take considerably longer time to process packets.

However, for the core switches the switching performance remains unaffected
compared to normal operation. This is because the core switches do
not perform any memory lookup when processing packets.

\subsection{Path Validation Overhead}
Path validation introduces processing overhead for the controller and bandwidth
overhead for the network. The controller has to recompute the PVFs for the
reported packets, and the egress switches have to report the PVFs in the packet
headers to the controller.

We estimate the overheads based on information for large data centers: 80k
hosts~\cite{dc-size} with 10G access links that are utilized at 1\% (in each
direction)~\cite{facebook-datacenter} and send average-sized packets of 850
bytes~\cite{data_traffic}. Due to lack of knowledge for traffic patterns,
we consider the worst case: all traffic is inter-rack and the path consists
of 5 switches (worst-case path length for 3-tier data center); also, all 
egress switches report all the packet headers. Overall, the aggregate packet
rate for this setup is 1176~Mpps.

We implement the PVC, which reads the packet headers, fetches the corresponding
shared keys with the switches, and recomputes the PVFs. For the previous setup, 
an 8 core CPU can validate 17~Mpps. For the whole data-center traffic
(1176~Mpps), 69 CPUs would be required.

For the bandwidth overhead, the data size to validate one packet is 14 bytes (3
bytes for the FlowID, 3 bytes for the SeqNo, and 8 bytes for the PVF).  For the
previous setup, we estimate the bandwidth overhead at 115~Gbps, which accounts
for 1.6\% of the whole data-center traffic.

\section{Discussion}
\label{sec:discussion}
\label{subsec:sec_con}
One attack we have not considered is packet dropping by a malicious
switch. Flow statistics through monitoring provide a basic
defense perimeter for such attacks. The controller can instruct switches
to periodically report packet counters for certain flows and then inspect
if packets are dropped at a certain link. Furthermore, dishonest reports 
would result in inconsistent reports that pinpoint the misbehavior to a
certain link between two switches (it is not possible to identify the exact
switch)~\cite{pqm}. However, packet dropping from a malicious ingress or
egress switch cannot be detected through monitoring. This is the side-effect
of a design decision in \name.

We have made the deliberate design decision that the network stack of the host
should not be modified. This design choice provides a smoother incremental
deployment path for \name, since hosts do not perform any additional
functionality.  This can be beneficial also for a data-center deployment, when
tenants have control over their operating system (\eg in the
Infrastructure-as-a-Service model).

This design decision, however, has implications for the security properties of
\name and enables certain attacks. For example, a malicious egress switch can
transfer packets out of an incorrect interface, replay packets, or drop
packets; without feedback from the end host it is not possible to detect such
attacks.  Furthermore, a malicious ingress switch can replay packets without
being detected, since the ingress switch can inscribe different sequence
numbers; again, the transport layer of the destination host -- and not the
network -- can detect the replay.

\section{Related Work}
\label{sec:related}
We briefly describe recent research proposals that are related to data-plane security
and state reduction in SDN.

\paragraph{Data-plane security} There are only a few proposals accounting for
compromised switches at the data plane.  The most closely related work to ours
is SANE~\cite{sane}: the controller hands out capabilities to end hosts -- not
to switches, as in \name~-- in order to enforce network paths.  This approach
requires modification of end hosts in order to perform additional tasks.
Namely, every host must communicate with the controller in order to establish a
shared symmetric key and obtain capabilities. Failure recovery is pushed to the
host, which has to detect the failure and then explicitly ask the controller
for a new path. In addition, SANE cannot provide protection against stronger
adversaries that collude and perform a wormhole attack: a malicious switch can
replay capabilities by prepending them to the existing forwarding information
in the packet and thus can diverge traffic over another path; a colluding
switch removes the prepended capabilities and forwards packets to a downstream
switch of the original path.  \name provides path validation to deal with such
attacks. Finaly, SANE does not consider broadcast/multicast forwarding.

Jacquin \etal \cite{switch_measurement} take another approach, using trusted computing to attest remotely
that network elements use an approved software version. Being a first step in this
direction, there are unaddressed challenges with respect to scalability: processing overhead
(overall attestation time), bandwidth overhead (extra traffic due to attestation), and
management overhead (the number of different software versions deployed).

OPT~\cite{opt} provides path validation on top of a dynamic key-establishment
protocol that enables routers to re-create symmetric keys with end hosts.
In \name, key management is simplified, since each router shares a key only with the controller.
Thus, we do not involve the host in any key establishment and avoid
the overhead of key establishment in the presented protocols.

ICING~\cite{icing} is another path validation protocol that leverages
cryptographic information in the packets. Each router on the path verifies
cryptographic markings in the packet that were inserted by the source and each
upstream router. ICING comes with a high bandwidth overhead due to large packet
sizes, demonstrating a 23.3\% average packet overhead. Furthermore, ICING
requires pairwise symmetric keys between all entities on a path.

\paragraph{State reduction for SDN}
Another class of proposals focuses on \textit{state reduction} for the SDN data
plane. Source routing is a commonly used approach to realize this goal, and
recent work shows that source routing not only decreases the forwarding table
size, but provides a higher and more flexible resource utiliziation~\cite{src_rout_dtc}.
In SourceFlow~\cite{Chiba:2010:SFH:1851182.1851266}, packets carry
pointers to action lists for every core switch on the path. Hence, core
switches only store action tables that encode potential actions for packets and
are indexed by a pointer in the packet. Segment Routing~\cite{segment_routing}
is based on source routing and combines the benefits of MPLS~\cite{rfc3031}
with the centralized control of SDN. An ingress switch adds an ordered list
of instructions into the packet header, and each subsequent switch inspects such
an instruction.  These approaches are similar to \name in that they reduce
state at core switches by embedding information in packet headers. However, the
use of source routing without corresponding security mechanisms opens a bigger
attack vector compared to legacy hop-by-hop routing: a single compromised
switch can modify the forwarding information and steer a packet over a
non-compliant path.

\section{Conclusion}
\label{sec:conclusion}
Security in SDN remains a neglected issue and could raise deployment hurdles
for security concerned environments.  We have presented a security extension to
achieve forwarding accountability for the SDN data plane, \ie to ensure that
the operator's policies are correctly applied to the data plane. To this end,
we have designed two mechanisms: path enforcement to ensure that the switches
forward the packets based on the instructions of the operator and path
validation to allow the operator to reactively verify that the data plane has
followed the specified policies. In addition, \name guarantees consistent
policy updates such that the behavior of the data plane is well defined during
reconfigurations.  Lastly, minimizing the amount of state at the core switches
confines state exhaustion attacks to the network edge.  We hope that this work
assists in moving towards more secure SDN deployments.

{
\bibliographystyle{myIEEEtran}
\bibliography{fullbib,rfc}

\begin{thebibliography}{10}
\providecommand{\url}[1]{#1}
\csname url@samestyle\endcsname
\providecommand{\newblock}{\relax}
\providecommand{\bibinfo}[2]{#2}
\providecommand{\BIBentrySTDinterwordspacing}{\spaceskip=0pt\relax}
\providecommand{\BIBentryALTinterwordstretchfactor}{4}
\providecommand{\BIBentryALTinterwordspacing}{\spaceskip=\fontdimen2\font plus
\BIBentryALTinterwordstretchfactor\fontdimen3\font minus
  \fontdimen4\font\relax}
\providecommand{\BIBforeignlanguage}[2]{{\expandafter\ifx\csname l@#1\endcsname\relax
\typeout{** WARNING: IEEEtran.bst: No hyphenation pattern has been}\typeout{** loaded for the language `#1'. Using the pattern for}\typeout{** the default language instead.}\else
\language=\csname l@#1\endcsname
\fi
#2}}
\providecommand{\BIBdecl}{\relax}
\BIBdecl

\bibitem{openflow}
\BIBentryALTinterwordspacing
N.~McKeown, T.~Anderson, H.~Balakrishnan, G.~Parulkar, L.~Peterson, J.~Rexford,
  S.~Shenker, and J.~Turner, ``Open{F}low: {E}nabling {I}nnovation in {C}ampus
  {N}etworks,'' \emph{SIGCOMM Comput. Commun. Rev.}, 2008.
\BIBentrySTDinterwordspacing

\bibitem{fresco}
\BIBentryALTinterwordspacing
S.~Shin, P.~A. Porras, V.~Yegneswaran, M.~W. Fong, G.~Gu, and M.~Tyson,
  ``{FRESCO: Modular Composable Security Services for Software-Defined
  Networks.}'' in \emph{Proc. of NDSS}, 2013.
\BIBentrySTDinterwordspacing

\bibitem{veriflow}
\BIBentryALTinterwordspacing
A.~Khurshid, W.~Zhou, M.~Caesar, and P.~B. Godfrey, ``{VeriFlow: Verifying
  Network-wide Invariants in Real Time},'' in \emph{Proc. of ACM HotSDN}, 2012.
\BIBentrySTDinterwordspacing

\bibitem{fortnox}
\BIBentryALTinterwordspacing
P.~Porras, S.~Shin, V.~Yegneswaran, M.~Fong, M.~Tyson, and G.~Gu, ``{A Security
  Enforcement Kernel for OpenFlow Networks},'' in \emph{Proc. of HotSDN}, 2012.
\BIBentrySTDinterwordspacing

\bibitem{syn_knock}
``{Cisco Routers Compromised by Malicious Code Injection},''
  "\url{http://bit.ly/1KtUoTs}", Sep. 2015.

\bibitem{juniper_bug}
``{Juniper ScreenOS Authentication Backdoor},'' "\url{http://bit.ly/1Nx8J5i}",
  Dec. 2015.

\bibitem{nsa_backdoor}
``{Snowden: The NSA planted backdoors in Cisco products},''
  "\url{http://bit.ly/1PKtbQW}", May 2015.

\bibitem{detectswitch}
P.-W. Chi, C.-T. Kuo, J.-W. Guo, and C.-L. Lei, ``{How to Detect a Compromised
  SDN Switch},'' in \emph{Proc. of IEEE NetSoft}, 2015.

\bibitem{kloti2013}
R.~Kloti, V.~Kotronis, and P.~Smith, ``{OpenFlow: A Security Analysis},'' in
  \emph{Proc. of IEEE NPSec}, 2013.

\bibitem{spooknet}
\BIBentryALTinterwordspacing
M.~Antikainen, T.~Aura, and M.~S\"arel\"a,
  ``\BIBforeignlanguage{English}{{Spook in Your Network: Attacking an SDN with
  a Compromised OpenFlow Switch}},'' in
  \emph{\BIBforeignlanguage{English}{Secure IT Systems}}, 2014.
\BIBentrySTDinterwordspacing

\bibitem{netupdate}
\BIBentryALTinterwordspacing
M.~Reitblatt, N.~Foster, J.~Rexford, C.~Schlesinger, and D.~Walker,
  ``{Abstractions for Network Update},'' in \emph{Proc. of ACM SIGCOMM}, 2012.
\BIBentrySTDinterwordspacing

\bibitem{sane}
M.~Casado, T.~Garfinkel, A.~Akella, M.~J. Freedman, D.~Boneh, N.~McKeown, and
  S.~Shenker, ``{SANE: A Protection Architecture for Enterprise Networks},'' in
  \emph{Proc. of USENIX Security}, Aug 2006.

\bibitem{hsa}
\BIBentryALTinterwordspacing
P.~Kazemian, G.~Varghese, and N.~McKeown, ``{Header Space Analysis: Static
  Checking for Networks},'' in \emph{Proc. of USENIX NSDI}, 2012.
\BIBentrySTDinterwordspacing

\bibitem{netplumber}
\BIBentryALTinterwordspacing
P.~Kazemian, M.~Chang, H.~Zeng, G.~Varghese, N.~McKeown, and S.~Whyte, ``{Real
  Time Network Policy Checking Using Header Space Analysis},'' in \emph{Proc.
  of USENIX NSDI}, 2013.
\BIBentrySTDinterwordspacing

\bibitem{anteater}
\BIBentryALTinterwordspacing
H.~Mai, A.~Khurshid, R.~Agarwal, M.~Caesar, P.~B. Godfrey, and S.~T. King,
  ``Debugging the data plane with anteater,'' in \emph{Proc. of ACM SIGCOMM},
  2011.
\BIBentrySTDinterwordspacing

\bibitem{8021X}
``{IEEE Standard for Local and metropolitan area networks - Port-Based Network
  Access Control},'' \emph{IEEE Std 802.1X-2004}, Dec 2004.

\bibitem{Mahajan:2013}
R.~Mahajan and R.~Wattenhofer, ``{On Consistent Updates in Software Defined
  Networks},'' in \emph{Proc. of ACM HotNets}, 2013.

\bibitem{Dionysus}
\BIBentryALTinterwordspacing
X.~Jin, H.~H. Liu, R.~Gandhi, S.~Kandula, R.~Mahajan, M.~Zhang, J.~Rexford, and
  R.~Wattenhofer, ``{Dynamic Scheduling of Network Updates},'' in \emph{Proc.
  of ACM SIGCOMM}, 2014.
\BIBentrySTDinterwordspacing

\bibitem{of15}
O.~N. Foundation, ``{OpenFlow Switch Specification Version 1.5.0},''
  "\url{http://bit.ly/1Rdi6Yg}", 2014.

\bibitem{datac_traffic_ccr}
\BIBentryALTinterwordspacing
T.~Benson, A.~Anand, A.~Akella, and M.~Zhang, ``{Understanding Data Center
  Traffic Characteristics},'' \emph{SIGCOMM Comput. Commun. Rev.}, 2010.
\BIBentrySTDinterwordspacing

\bibitem{datac_traffic}
\BIBentryALTinterwordspacing
T.~Benson, A.~Akella, and D.~A. Maltz, ``{Network Traffic Characteristics of
  Data Centers in the Wild},'' in \emph{Proc. of ACM IMC}, 2010.
\BIBentrySTDinterwordspacing

\bibitem{lipsin}
P.~Jokela, A.~Zahemszky, C.~Esteve~Rothenberg, S.~Arianfar, and P.~Nikander,
  ``{LIPSIN: Line Speed Publish/Subscribe Inter-networking},'' in \emph{Proc.
  of ACM SIGCOMM}, 2009.

\bibitem{mech_pathv}
\BIBentryALTinterwordspacing
F.~Zhang, L.~Jia, C.~Basescu, T.~H.-J. Kim, Y.-C. Hu, and A.~Perrig,
  ``{Mechanized Network Origin and Path Authenticity Proofs},'' in \emph{Proc.
  of ACM CCS}, 2014.
\BIBentrySTDinterwordspacing

\bibitem{dpdk}
``Data {P}lane {D}evelopment {K}it,'' "\url{http://dpdk.org}".

\bibitem{aesni}
S.~Gueron, ``{Intel Advanced Encryption Standard (AES) New Instructions Set
  },'' 2012.

\bibitem{rss}
\BIBentryALTinterwordspacing
S.~Goglin and L.~Cornett, ``Flexible and extensible receive side scaling,''
  2009.
\BIBentrySTDinterwordspacing

\bibitem{dpdkvswitch}
``{Open vSwitch accelerated by DPDK},''
  "\url{https://github.com/01org/dpdk-ovs}".

\bibitem{openvswitch}
``{Open vSwitch},'' "\url{www.openvswitch.org}".

\bibitem{leafspine}
\BIBentryALTinterwordspacing
M.~Alizadeh and T.~Edsall, ``{On the Data Path Performance of Leaf-Spine
  Datacenter Fabrics},'' in \emph{Proc. of IEEE HOTI}, 2013.
\BIBentrySTDinterwordspacing

\bibitem{multitier}
Cisco, ``{Data Center Multi-Tier Model Design},''
  "\url{http://bit.ly/23vtt5u}".

\bibitem{internet2}
``{Internet2},'' "\url{https://www.internet2.edu}".

\bibitem{internet2_noc}
``{Internet2 Network NOC},'' "\url{http://bit.ly/1JHulh0}".

\bibitem{caida}
``{CAIDA}: {C}enter for {A}pplied {I}nternet {D}ata {A}nalysis,''
  "\url{http://www.caida.org}".

\bibitem{jellyfish}
\BIBentryALTinterwordspacing
A.~Singla, C.-Y. Hong, L.~Popa, and P.~B. Godfrey, ``{Jellyfish: Networking
  Data Centers Randomly},'' in \emph{Proc. of USENIX NSDI}, 2012.
\BIBentrySTDinterwordspacing

\bibitem{dc-size}
``{Inside Amazon’s Cloud Computing Infrastructure},''
  "\url{http://bit.ly/1JHulh0}", 2015.

\bibitem{facebook-datacenter}
A.~Roy, H.~Zeng, J.~Bagga, G.~Porter, and A.~C. Snoeren, ``Inside the social
  network's (datacenter) network,'' in \emph{Proceedings of the 2015 ACM
  Conference on Special Interest Group on Data Communication}.

\bibitem{pqm}
\BIBentryALTinterwordspacing
S.~Goldberg, D.~Xiao, E.~Tromer, B.~Barak, and J.~Rexford, ``Path-quality
  {M}onitoring in the {P}resence of {A}dversaries,'' in \emph{Proc. of ACM
  SIGMETRICS}, 2008.
\BIBentrySTDinterwordspacing

\bibitem{switch_measurement}
L.~Jacquin, A.~Shaw, and C.~Dalton, ``{Towards trusted software-defined
  networks using a hardware-based Integrity Measurement Architecture},'' in
  \emph{Proc. of IEEE NetSoft}, 2015.

\bibitem{opt}
\BIBentryALTinterwordspacing
T.~H.-J. Kim, C.~Basescu, L.~Jia, S.~B. Lee, Y.-C. Hu, and A.~Perrig,
  ``Lightweight {S}ource {A}uthentication and {P}ath {V}alidation,'' in
  \emph{Proc. of ACM SIGCOMM}, 2014.
\BIBentrySTDinterwordspacing

\bibitem{icing}
\BIBentryALTinterwordspacing
J.~Naous, M.~Walfish, A.~Nicolosi, D.~Mazi\`{e}res, M.~Miller, and A.~Seehra,
  ``Verifying and {E}nforcing {N}etwork {P}aths with {ICING},'' in \emph{Proc.
  of ACM CoNEXT}, 2011.
\BIBentrySTDinterwordspacing

\bibitem{src_rout_dtc}
\BIBentryALTinterwordspacing
S.~A. Jyothi, M.~Dong, and P.~B. Godfrey, ``{Towards a Flexible Data Center
  Fabric with Source Routing},'' in \emph{Proc. of ACM SOSR}, 2015.
\BIBentrySTDinterwordspacing

\bibitem{Chiba:2010:SFH:1851182.1851266}
Y.~Chiba, Y.~Shinohara, and H.~Shimonishi, ``{Source Flow: Handling Millions of
  Flows on Flow-based Nodes},'' in \emph{Proc. of ACM SIGCOMM}, 2010.

\bibitem{segment_routing}
``{Segment Routing Architecture},''
  "\url{https://tools.ietf.org/html/draft-filsfils-rtgwg-segment-routing-01"}.

\bibitem{rfc3031}
\BIBentryALTinterwordspacing
E.~Rosen, A.~Viswanathan, and R.~Callon, ``{Multiprotocol Label Switching
  Architecture},'' RFC 3031 (Proposed Standard), Internet Engineering Task
  Force, Jan. 2001.
\BIBentrySTDinterwordspacing

\end{thebibliography}
}

\end{document}